\documentclass[11pt]{article}
\usepackage{moriond,epsfig}
\usepackage{amssymb}

\def\be{\begin{equation}}
\def\ee{\end{equation}}
\def\bea{\begin{eqnarray}}
\def\eea{\end{eqnarray}}

\begin{document}
\vspace*{4cm}
\title{LONGITUDINAL DOUBLE  SPIN  ASYMMETRY IN JET PRODUCTION AT  STAR}

\author{ Katarzyna Kowalik for the STAR Collaboration}

\address{Lawrence Berkeley National Laboratory, Berkeley, California 94720 }

\maketitle\abstracts{ We report measurements of the longitudinal
double-spin asymmetry $A_{LL}$ for the inclusive production of jets at
midrapidity in polarized proton-proton collisions at
$\sqrt{s}=200\,\mathrm{GeV}$.  The data amount to an integrated
luminosity of $3\,\mathrm{pb}^{-1}$ and were collected with the STAR
detector at the Relativistic Heavy Ion Collider. Typical beam
polarizations were 50\%.  The $A_{LL}$ measurements cover jet
transverse momenta $5 < p_{T} < 30\,\mathrm{GeV/c}$ and 
provide sensitive constraints on the gluon spin contribution to the
proton spin.  }

\section{Introduction}
The Relativistic Heavy Ion Collider (RHIC) is the first polarized
 high-energy proton-proton collider, providing collisions at $\sqrt{s}
 =200\,\mathrm{GeV}$ and in the future at $\sqrt{s}
 =500\,\mathrm{GeV}$.  One of the main objectives of the program with
 polarized proton beams is the measurement of the gluon spin
 contribution, $\Delta G$, to the proton spin~\cite{Bunce:2000uv}.  The
 processes under study include inclusive jet and pion production,
 dijets, heavy flavor, and photon-jet coincidences.  The present focus
 is on inclusive measurements with large production cross sections.
 At a center-of-mass energy $\sqrt{s} =200\,\mathrm{GeV}$ production processes
 at midrapidity and with transverse momenta $p_T > 5\,\mathrm{GeV/c}$ 
typically give sensitivity to the integral of $\Delta G$ over the
 range $0.03 \lesssim x \lesssim 0.3$ in the gluon momentum fraction.
Coincidence measurements can provide the $x$ dependence of $\Delta
 G$.  However, they require larger total integrated luminosities than
 have been sampled so far.  Collisions at $\sqrt{s} =500\,\mathrm{GeV}$ will
 extend the x coverage to smaller values, which are important for
 global analyses of the polarized parton distributions.
 The STAR (Solenoidal Tracker at RHIC)
 detector~\cite{star}, with its large acceptance and electromagnetic
 calorimetry, is well suited for jet reconstruction at RHIC.
 
 The (polarized) jet cross section has large
 contributions from (polarized) gluon-gluon and  quark-gluon scattering,
which provide direct sensitivity to $\Delta G$ in the measurements of
 the longitudinal double-spin asymmetry $A_{LL}$,
\begin{equation}
  A_{LL} = \frac{\sigma^{++} - \sigma^{+-}}{\sigma^{++} + \sigma^{+-}},
\end{equation}
where $\sigma^{++}$ ($\sigma^{+-}$) is the inclusive jet cross section
when the colliding proton beams have equal (opposite) helicities.  The
gluon-gluon scattering contribution dominates up to jet $p_T
\lesssim 8\,\mathrm{GeV/c}$ while at higher $p_T$ in the range of the
present measurements the quark-gluon scattering contribution is the
largest. The quark-quark  contribution is small~\cite{Jager:2004jh}. 

First inclusive jet $A_{LL}$ and cross section results from STAR have
been published~\cite{Abelev:2006uq}.  The unpolarized cross section is well described
by NLO pQCD calculations for jet $p_T$, $5 < p_T <
50\,\mathrm{GeV/c}$, motivating the application of the NLO pQCD
framework to interpret the spin asymmetry results. The asymmetry
$A_{LL}$ was measured for $5 < p_{T} < 17\,\mathrm{GeV/c}$ and
disfavors large positive gluon polarization in the proton.  These
proceedings report on new measurements of $A_{LL}$ for inclusive jet
production with increased sensitivity and extended coverage in jet
$p_T$.

\section{Experiment and Data Analysis}

The present results are based on an integrated luminosity of 
$\sim 3\,\mathrm{pb}^{-1}$ and were recorded in the year 2005.  The STAR
detector subsystems used for jet reconstruction are the Time
Projection Chamber (TPC) and the Barrel Electromagnetic Calorimeter
(BEMC).  The TPC provides the momentum of charged particles in the
pseudorapidity range $-1.3 \lesssim \eta \lesssim 1.3$ for all
azimuthal angles~$\phi$. The BEMC is a lead-scintillator sampling
calorimeter which measures electromagnetic energy deposits. During the
data taking period in 2005, the BEMC covered $0 < \eta < 1$ and
$0<\phi<2\pi$. The remaining half, covering $-1 < \eta < 0$, was
commissioned before the data taking in 2006.  
Beam-Beam Counters (BBC),
located on each side of the STAR interaction region, provided the beam
collision trigger and were used to measure the relative luminosities
for different helicity configurations~\cite{Kiryluk:2005gg} as well as
transverse beam polarization components at STAR.

The majority of the jet data were collected using a new dedicated
jet-patch (JP) trigger that required a transverse energy sum in at
least one of six BEMC patches, each covering $\Delta\eta \times
\Delta\phi = 1 \times 1$. The JP trigger efficiency is higher than the
efficiency of the high tower trigger (HT), which selects on an energy
deposit in a BEMC tower of size $\Delta{\eta} \times \Delta{\phi}=0.05
\times 0.05$ and was used also in previous data taking periods.  In
addition, the JP trigger selects a less biased distribution of jets
than the HT trigger does.The HT and JP triggers were used with two
different energy thresholds.

Jets were reconstructed with a midpoint cone
algorithm~\cite{Blazey:2000qt} with a cone radius of 0.4 and using
charged TPC tracks and BEMC energy deposits. The details for other
parameters can be found in Ref.~\cite{Abelev:2006uq}.  Only jets with
reconstructed transverse momenta $p_T >5\,\mathrm{GeV/c}$ that
fulfilled the trigger conditions were considered in the analysis.  
Jets with their axis between a nominal $\eta$ of 0.2 and 0.8 were
selected so that the effects from the BEMC acceptance edges were
small.  BBC timing information was used to accept events with vertex positions along the beam direction 
in the inner region of STAR for which tracking efficiencies are uniform.
The same timing information was used in the beam luminosity measurement.
Beam background caused an occasional signal in the BEMC.  Its
contribution to the jet yield was suppressed by requiring the ratio of
jet energy in the BEMC to the total jet energy to be between 0.1 and
0.8.

\section{Results}

The jet yields were sorted by  equal  ($N^{++}$)   and opposite ($N^{+-}$)  helicity
combinations of the colliding proton  beams and $A_{LL}$ was evaluated according to:
\begin{equation}
  A_{LL}=\frac{ \sum (P_1 P_2) (N^{++}-R N^{+-}) }  {  \sum (P_1 P_2)^2 ( N^{++}+ R N^{+-})},
\end{equation} 
where $P_1 P_2$ is the product of the measured beam
polarizations~\cite{Jinnouchi:2004up,Okada:2006dd}, and $R$ is the
measured ratio of luminosities for equal and opposite proton beam
helicities~\cite{Kiryluk:2003aw}. The average online beam polarization
was $\sim\,$50\%. The ratio $R$ was between 0.8 and 1.2, and was
measured to $\mathcal{O}{(10^{-3})}$ accuracy.  The yields $N^{++}$ and
$N^{+-}$ were recorded concurrently since the proton beam helicities
alternated for successive beam bunches in one beam and for successive
pairs of beam bunches in the other beam.  To further minimize
systematic effects in the measurement of $A_{LL}$, the beam helicity
pattern was alternated between RHIC beam fills.

The asymmetry was calculated for the combined sample of about 1.97\,M
jets of JP and HT triggered data.  
Figure~\ref{fig:all} shows $A_{LL}$ as a function of jet $p_T$ with
statistical error bars for the present and published~\cite{Abelev:2006uq} data and
the systematic uncertainty band for the present data, not including
the 25\% scale uncertainty from the online beam polarization
measurement.  The leading contributions to the systematic uncertainty
band arise from the bias introduced by the BEMC trigger requirements
and from a conservative upper limit on possible false asymmetries in
the measurement.  Other systematic uncertainties include effects from
non-longitudinal beam polarization components at the STAR interaction
region, from uncertainty in the measurement of $R$, and from beam
background.  Systematic checks with randomized beam-spin patterns
showed no evidence for bunch-to-bunch or fill-to-fill systematics in
$A_{LL}$.
 
The curves in Figure~\ref{fig:all} show NLO pQCD evaluations for
$A_{LL}$ based on commonly used polarized parton distribution
functions~\cite{Jager:2004jh,Gluck:2000dy}.  The curve labeled
GRSV-std is based on a best fit to inclusive DIS
data~\cite{Gluck:2000dy}.  The other curves correspond to maximally
positive ($\Delta g=g$), negative ($\Delta g=-g$), or vanishing
($\Delta g=0$), gluon polarizations at a $0.4\,\mathrm{GeV^2/c^2}$
initial scale of the parton parametrizations~\cite{Gluck:2000dy}.

\begin{figure}
\label{fig:all}
\begin{center}
\psfig{figure=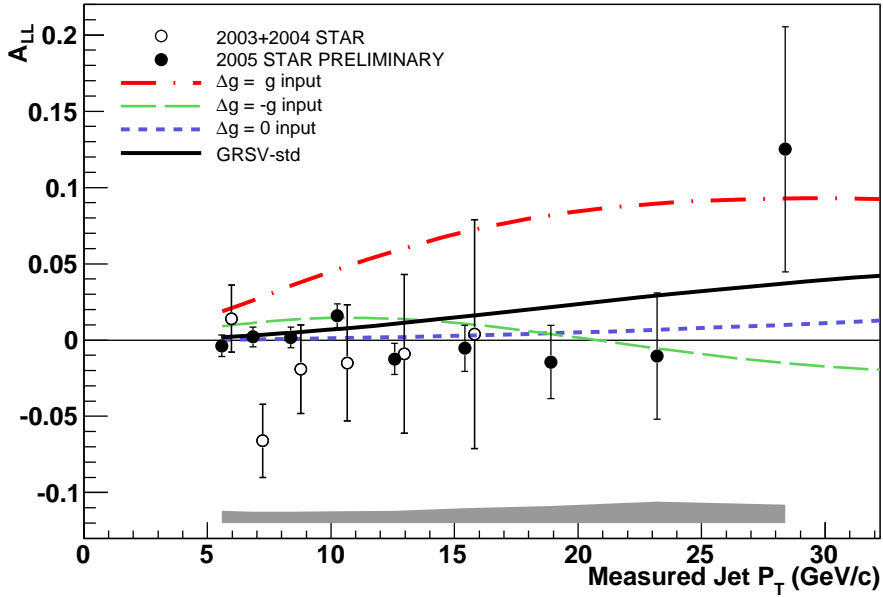,height=3.5in}
\end{center}
\caption{The longitudinal double-spin asymmetry $A_{LL}$ versus jet
transverse momentum $p_T$ for inclusive jet production in polarized
proton-proton collisions at $\sqrt{s}=200\,\mathrm{GeV}$.  The filled
symbols show the present data and the open symbols the published
results.  The error bars indicate the size of the statistical
uncertainties. The band shows the size of the total systematic
uncertainty in the present data. The uncertainty from the beam
polarization is not included.  The curves show different theoretical
predictions for gluon polarization in the polarized proton and are
discussed in the text.
\label{fig:all}
}
\end{figure}

The present data  are in good agreement with the our published 
results~\cite{Abelev:2006uq} in the region of kinematic overlap $5 <
p_T <17 \,\mathrm{GeV/c}$, where they improve significantly the precision, 
and extend jet  $p_T$  up to $30\,\mathrm{GeV/c}$.
The corresponding range of gluon momentum fractions  sampled by data
is  $0.03  \lesssim x  \lesssim 0.3$. The fraction of the first moment of
the GRSV-std polarized gluon distribution is about  half  over this $x$ range.
The data are not consistent with the GRSV  scenario of maximal positive 
gluon spin contribution to the proton spin ($\Delta g=g$).

An additional $8.5\,\mathrm{pb}^{-1}$ was sampled in 2006 with beam
polarizations of $\sim\,$60\%. This will improve the precision of
inclusive $A_{LL}$ measurements and offers good prospects for dijet analyses.

\section{Summary}
We reported preliminary measurements of the longitudinal double-spin
asymmetry $A_{LL}$ for the inclusive production of jets in polarized
proton-proton collisions at $\sqrt{s}=200\,\mathrm{GeV}$ with jet
$p_T$ up to $30\,\mathrm{GeV/c}$.  The $A_{LL}$ data, compared to the
commonly used GRSV set of polarized parton distributions, exclude the
scenario with a large positive gluon spin contribution to the nucleon
spin.  Future inclusion of our data as well as data in
Ref.~\cite{dis,phenix} in global analyses should improve our knowledge
of the polarized gluon distribution for $0.03 \lesssim x \lesssim
0.3$. First promising works in this direction have already  been published~\cite{Hirai:2006sr}. 
\\ \\ {\bf{Acknowledgments}} \\ It is a pleasure to thank
organizers for a stimulating conference and the European Union "Marie Curie" Programme for
partial support.


\section*{References}
\begin{thebibliography}{99}

\bibitem{Bunce:2000uv}
  G.~Bunce, N.~Saito, J.~Soffer and W.~Vogelsang,
  Ann.\ Rev.\ Nucl.\ Part.\ Sci.\  {\bf 50} (2000) 525.

\bibitem{star} Special Issue: RHIC and Its Detectors,
\emph{Nucl. Instrum. Meth.} {\bf{A499}}, (2003).


\bibitem{Jager:2004jh}
  B.~J\"{a}ger, M.~Stratmann and W.~Vogelsang,
  Phys.\ Rev.\  D {\bf 70}, 034010 (2004).

\bibitem{Abelev:2006uq}
  B.~I.~Abelev {\it et al.}  [STAR Collaboration],
  Phys.\ Rev.\ Lett.\  {\bf 97}, 252001 (2006).


\bibitem{Kiryluk:2005gg}
  J.~Kiryluk  [STAR Collaboration],
  arXiv:hep-ex/0501072. 




\bibitem{Blazey:2000qt}
  G.~C.~Blazey {\it et al.},
  arXiv:hep-ex/0005012.





\bibitem{Jinnouchi:2004up}
  O.~Jinnouchi {\it et al.},
  arXiv:nucl-ex/0412053.

\bibitem{Okada:2006dd}
  H.~Okada {\it et al.},
  arXiv:hep-ex/0601001.

\bibitem{Kiryluk:2003aw} J.~Kiryluk for STAR Collaboration, 
  AIP Conf.\ Proc.\ {\bf 675}, 424 (2003).



\bibitem{Gluck:2000dy}
  M.~Gl\"{u}ck, E.~Reya, M.~Stratmann and W.~Vogelsang,
  Phys.\ Rev.\  D {\bf 63}, 094005 (2001).



\bibitem{dis}
  HERMES, A.~Airapetian {\it et al.},
  Phys.\ Rev.\ Lett.\  {\bf 84}, 2584 (2000);
  SMC, B.~Adeva {\it et al.},
  Phys.\ Rev.\  D {\bf 70}, 012002 (2004);
  COMPASS, E.~S.~Ageev {\it et al.},
  Phys.\ Lett.\  B {\bf 633}, 25 (2006).






\bibitem{phenix} 
  PHENIX, S.~S.~Adler {\it et al.},
  Phys.\ Rev.\  D {\bf 73}, 091102 (2006);
  arXiv:0704.3599 [hep-ex].




\bibitem{Hirai:2006sr}
  M.~Hirai, S.~Kumano and N.~Saito,
  Phys.\ Rev.\  D {\bf 74}, 014015 (2006);
  G.~A.~Navarro and R.~Sassot,
  Phys.\ Rev.\  D {\bf 74}, 011502 (2006).

\end{thebibliography}
\end{document}